\documentclass{epl}

\newcommand{\be}{\begin{equation}}
\newcommand{\ee}{\end{equation}}
\newcommand{\bea}{\begin{eqnarray}}
\newcommand{\eea}{\end{eqnarray}}

\newcommand{\Ca}{\mbox{Ca}}

\title{Slip-controlled thin film dynamics}
\author{R. Fetzer\inst{1} \and M. Rauscher\inst{2}\inst{3} \and A. M\"unch\inst{4}\inst{5} \and B. A. Wagner\inst{5} \and K. Jacobs\inst{1}
}

\institute{ \inst{1} Department of Experimental Physics, Saarland
University, 66123 Saarbr\"ucken, Germany\\ \inst{2} Max Planck
Institute for Metals Research, Heisenbergstra{\ss}e 3, 70569
Stuttgart, Germany\\ \inst{3} Institute for Theoretical and
Applied Physics, University of Stuttgart, Pfaffenwaldring 75,
70569 Stuttgart, Germany\\ \inst{4} Institute of Mathematics,
Humboldt University Berlin, 10099 Berlin, Germany \\ \inst{5}
Weierstrass Institute for Applied Analysis and Stochastics,
Mohrenstra{\ss}e 39, 10117 Berlin, Germany}


\begin{document}

\maketitle

\begin{abstract}
In this study, we present a novel method to assess the slip length
and the viscosity of thin films of highly viscous Newtonian
liquids. We quantitatively analyse dewetting fronts of low
molecular weight polystyrene melts on Octadecyl- (OTS) and
Dodecyltrichlorosilane (DTS) polymer brushes. Using a thin film
(lubrication) model derived in the limit of large slip lengths, we
can extract slip length and viscosity. We study polymer films with
thicknesses between 50~nm and 230~nm and various temperatures
above the glass transition. We find slip lengths from 100~nm up to
1~$\mu$m on OTS and between 300~nm and 10~$\mu$m on DTS covered
silicon wafers. The slip length decreases with temperature. The
obtained values for the viscosity are consistent with independent
measurements.
\end{abstract}

\section{Introduction}
\label{intro} 
\begin{figure}
\centerline{
\includegraphics*[width=0.8\linewidth]{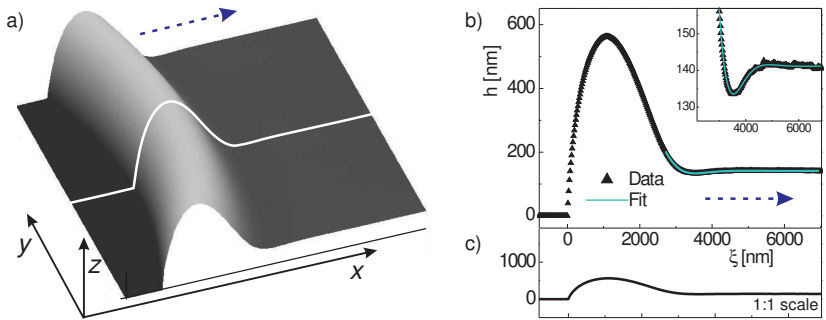}}
\caption{Profile of a moving front. a) SPM image of a section of
the rim around a hole in a 130~nm thick PS(13.7k) film on OTS at
120$^\circ$C, scan size 10~$\mu$m. The dashed arrow indicates the
direction of rim motion. b) A cross section taken in radial
direction (white line in the left image) gives the rim profile
(triangles) which is well fitted by an exponentially decaying
oscillation (solid line). The inset shows an enlarged view of the
oscillation. c) Data of b in 1:1 scale.} \label{fig1}
\end{figure}

Miniaturization of chemical appliances into
so-called microfluidic devices allows to handle smaller and
smaller amounts of liquid. However, in narrow channels, the effect
of the hydrodynamic boundary conditions becomes extremely
important. In particular, small amounts of slip on the channel
walls can improve throughput and decrease the dispersion of
chemical signals \cite{Son03}. In electronics industry,
downscaling photolitographic processes requires extremely thin
films of photoresist, a polymeric liquid. The dynamics of these
thin films is also significantly affected by the boundary
condition at the solid/liquid interface. Hence, there is a strong
need to quantify slippage of different liquids on various
substrates, and to characterise the influence of system parameters
on the effective slip length. Common techniques to measure the
velocity at the substrate involve tracer particles
\cite{Tre02,Tre04,Lum03} or fluorescence recovery after
photobleaching \cite{Pit00,Leg03,Sch05}. In addition, there are
various indirect methods to determine the amount of slippage,
mostly drainage experiments, e.g., in a surface forces apparatus
\cite{Cot02,Zhu02,Zhu02b} or between a colloidal probe particle
and a wall \cite{Cra01,Vin01,Vin03}. For recent reviews see
refs.~\cite{Net05,Lau06}. As none of these techniques can be
applied to all liquid/substrate-combinations, we present here an
alternative method which is most suited for highly viscous fluids
on partially and non-wetting substrates. We use the dewetting
process of supported polystyrene (PS) films to induce flow. Slip
length and capillary number are then extracted from the shape of
the rim around the growing holes in the dewetting film (see for
example \ref{fig1}). For determining in addition the viscosity of
the liquid with the help of our model, we independently measure
the rim velocity.

\section{Experiments}
\label{Exp} The liquids we used in our experiments were atactic
polystyrene (PS) melts with molecular weight 5.61~kg/mol,
13.7~kg/mol, and 18~kg/mol (each $M_n /M_w =1.06$, PSS Mainz,
Germany). We prepared polymer films with thickness 50(3)~nm,
130(5)~nm, and 230(5)~nm by spin casting from a toluene solution
onto mica, floating on Millipore water, and picking up by
hydrophobised Si wafers (Wacker, Burghausen, RMS roughness
0.1~nm). We used standard techniques to hydrophobise wafers with
self assembled Octadecyl- and Dodecyltrichlorosilane (OTS and DTS)
monolayers \cite{Was89}. The receeding contact angle of PS
droplets on OTS as well as on DTS covered Si wafers is
67(3)$^\circ$.

After heating the films above their glass transition temperature
(about 100$^\circ$C) dewetting takes place. Holes nucleate, grow,
and coalesce until in the later stage of dewetting only a set of
droplets remain on the substrate. During the early stage of this
process, we measured the radii of the radial growing circular
holes, $R$, as a function of time with optical microscopy to
determine the dewetting velocity. Once the holes had reached a
radius of about 12~$\mu$m, we immediately quenched our samples to
room temperature to bring the samples in the glassy state. Using
scanning probe microscopy (SPM Multimode, Digital Instruments,
Santa Barbara, USA) in Tapping Mode$^{TM}$ we measured the glassy
rim profiles. To ensure that the rim does not change its shape
while cooling we also scanned liquid profiles, but could not
detect any difference in shape.

\section{Thin film analysis}

Flow of thin Newtonian liquid films with no or weak slip at the
substrate is well described by the thin film equation as discussed
in detail, e.g., in ref.~\cite{Oro97}. The essence of the
underlying lubrication approximation is the assumption that the
lateral scale of thickness variations in the film $L$ is large as
compared to the characteristic film thickness $H$ (in this paper
the thickness of the undisturbed film). In the linear Navier slip
model, the amount of slip at the solid/liquid interface is
characterised by the slip length $b$. The boundary condition for
the velocity component parallel to the substrate is $u = b
\partial u/\partial z$. Recently, a thin film model has been developed for the
case where the slip length is much larger than the film thickness
scale $H$ \cite{Fet05,Kar04,Mue05}. Neglecting inertia, the
lateral velocity in the film for this model is given by

\begin{equation}
\label{eqvelo} u=\frac{2\,b}{h}\,\partial_x
\left(2\,\eta\,h\,\partial_x u\right) +
\frac{b\,h}{\eta}\partial_x\left(\sigma\,\partial^2_x h\right),
\end{equation}
with the viscosity  $\eta$  and the surface tension  $\gamma$\/.
In eq. (\ref{eqvelo}), we neglect the influence of the disjoining
pressure that arises due to the long-range intermolecular forces
which drive the dewetting, since we only study regions of the film
with a minimum thickness of about 50~nm. However, these forces can
be easily included in the model \cite{Mue05}. The first term on
the right side is proportional to the divergence of the total
longitudinal shear stress integrated over the film thickness. The
second term is the gradient of the Laplace pressure. The dynamics
of the film thickness induced by the flow is then obtained from
the continuity equation

\begin{equation}
\label{eqmcons}
\partial_t h + \partial_x (h\,u)=0.
\end{equation}

In order to analyse the decay of the rim profile close to the
undisturbed film of thickness $H$, see inset to fig.~\ref{fig1}b,
we linearise eqs. (\ref{eqvelo}) and (\ref{eqmcons}) about the
undisturbed state of thickness $h=H$ and $u=0$ by introducing a
small perturbation $\delta h(x,t) = h(x,t) - H$ and a small
velocity $u(x,t)$. The resulting linear equation for $\delta
h(x,t)$ is equivalent to the Newtonian limit of the model used in
ref.~\cite{Her02} to investigate the rim shape of viscoelastic
dewetting films.

We assume that the shape change of the rim due to accumulation of
liquid during hole growth is slow as compared to the relaxation
time for $\delta h(x,t)$ and $u(x,t)$. The experimental
observation indeed shows that dewetting is much faster than the
growth of the rim. Then, in a frame of reference  $\xi = x-s(t)$
comoving with the position of the rim $s(t)$ we have a
quasi-stationary profile. We solve the linear equation with the
normal modes ansatz $\delta h=\delta h_0 \exp\{\kappa \xi\}$ and
$u=u_0 \exp\{\kappa \xi \}$, and get the following characteristic
equation for $\kappa$

\begin{equation}
\label{eq_polynomial} (H\,\kappa)^3 + 4\,\Ca\,(H\,\kappa)^2 -
\Ca\,\frac{H}{b} = 0,
\end{equation}

with the capillary number $\Ca=\frac{\eta\,\dot{s}}{\sigma}$ and
the speed of the rim $\dot{s}= \partial_t s$\/.
eq.~(\ref{eq_polynomial}) is a polynomial of third order and has
therefore three solutions, one of which is real and positive and
therefore unphysical. For $\Ca^2 < 3^3\,H/(4^4\, b)$, i.e., for
slowly dewetting films or for moderately large slip length, the
remaining two solutions are a complex conjugate pair
$\kappa=\kappa_r\pm i\kappa_i$ with $\kappa_r<0$\/. In this case,
the rim is expected to have an oscillatory shape as shown in
fig.~\ref{fig1}. In fact, even for small slip lengths (where the
model used here is not valid) rim shapes are always oscillatory,
in particular in the no-slip case \cite{See01}. For rapidly
growing holes or large slip lengths both remaining solutions
$\kappa_1$ and $\kappa_2$ are real and negative, which allows rims
to decay monotonically. The aforementioned inequality also implies
that the transition in the rim shapes can occur even for fixed
$b$, if the film thickness $H$ is changed; increasing $H$ has the
same effect as decreasing $b$ (with the other quantity held
fixed).

\begin{figure}
\centerline{
\includegraphics*[width=0.9\linewidth]{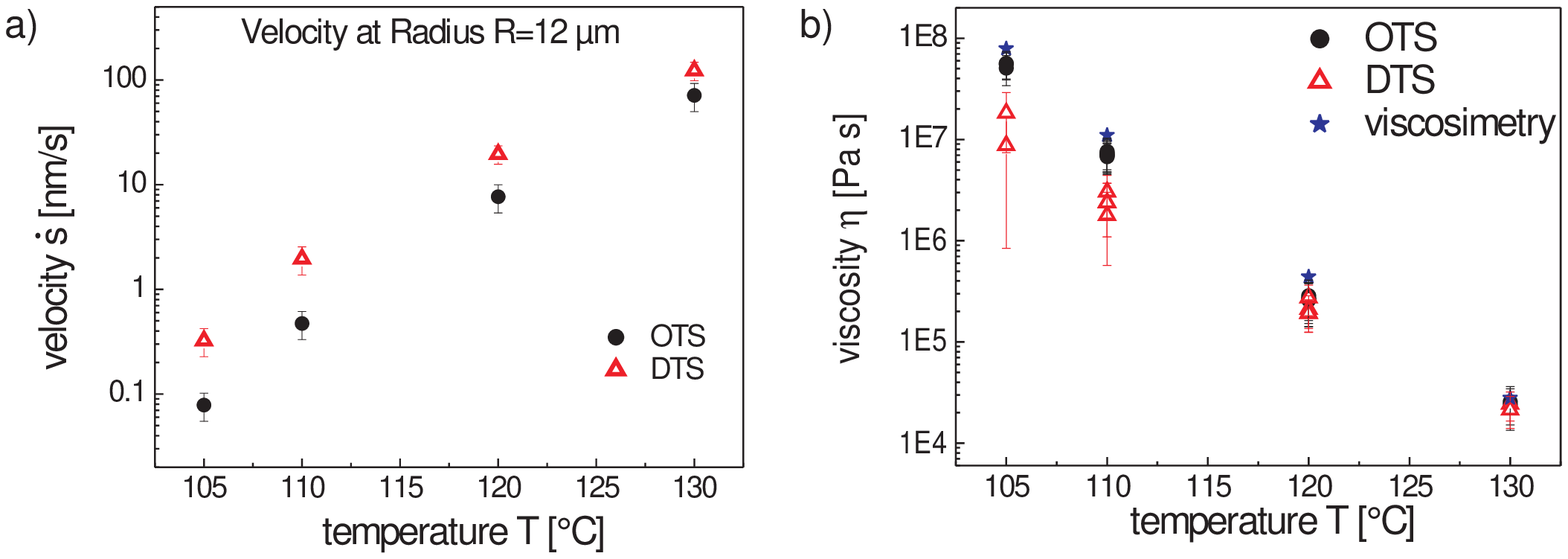}}
\caption{a) Rim velocity of holes with a radius of 12~$\mu$m in
130~nm thick PS(13.7k) films dewetting from OTS and DTS covered
silicon wafers as a function of temperature. The velocities were
measured by optical microscopy. Note the logarithmic velocity
scale. b) Viscosity as a function of temperature extracted from
rim profiles of 130~nm thick PS(13.7k) films which dewet from OTS
and DTS. As expected, the obtained values for the viscosity are
identical on both substrates and in good agreement with
independent viscosimetry data. \label{fig2}}
\end{figure}

With scanning probe microscopy (SPM) we measured the rim shape of
dewetting liquid PS(13.7k) films of 130~nm thickness on
hydrophobised Si wafers (for details see section "Experiments").
We also repeated some experiments with thinner (50~nm) and thicker
(230~nm) films, and longer (18k) as well as shorter (5.61k) chain
lengths. Depending on the type of substrate coating, densely
grafted Octadecyl- (OTS) or Dodecyltrichlorosilane (DTS) polymer
brushes, and on the dewetting temperature, we observed oscillatory
or monotonically decaying profiles. The observed contact angles of
polystyrene droplets which are left on DTS and OTS after the
dewetting process are both 67(3)$^\circ$, yet the rim velocities
on DTS are much larger than on OTS, cf. fig.~\ref{fig2}a),
indicating a difference in slip length. We like to emphasize,
however, that the method we develop here for determining the slip
length does not depend on the contact angle.

As demonstrated in fig.~\ref{fig1} for the case of an oscillatory
profile, an exponentially damped oscillation  $\delta h_{osci} =
\delta h_0 \exp(\kappa_r \xi) \cos(\kappa_i \xi + \phi )$ (fit
parameters are $\delta h_0$, $\kappa_i$, $\kappa_r$, and $\phi$)
captures the decay towards the resting film thickness in the
experimental data very well. From the fit we gain the inverse
decay length  $\kappa_r$ and the wave number $\kappa_i$, and from
independent measurements we know the surface tension $\sigma$ and
the film thickness $H$. Inserting  $\kappa = \kappa_r \pm i
\kappa_i$ into eq. (\ref{eq_polynomial}) and separating real and
imaginary part we get two linear equations for the two unknowns Ca
and $b$, which can easily be solved.

In the case of monotonically decaying rims, we fit the data with a
superposition of two exponentials $\delta h_{mono} = \delta h_1
\exp( \kappa_1 \xi )+ \delta h_2 \exp( \kappa_2 \xi )$ (fit
parameters $\delta h_{1/2}$ and $\kappa_{1/2}$) with inverse decay
lengths $\kappa_1$ and $\kappa_2$. Both, $\kappa_1$ and
$\kappa_2$, fulfill eq. (\ref{eq_polynomial}) such that we again
get a simple system of two linear equations for Ca and $b$.

Additionally, we can determine the film viscosity $\eta$  from the
capillary number Ca, using the surface tension  $\sigma =
30.8$~mN/m and the observed dewetting velocity $\dot{s}$. We like
to emphasize that in order to determine solely the slip length,
the knowledge of both the dewetting velocity and the viscosity is
not required.

\section{Results}

In order to test the consistency of our method, we performed
experiments with films of various thicknesses, and we analysed the
rim shapes of the same film at different times, i.e., at different
hole sizes and therefore at different rim velocities. The values
for the viscosity should only depend on temperature and on chain
length, whereas the slip length can additionally depend on the
type of substrate.

Fig.~\ref{fig2}b) shows the results for the viscosity gained from
qualitatively very different profiles of 130~nm thick films on OTS
(oscillatory) and DTS (monotonic). They are in very good
agreement. Additionally, they are in accordance with the values
obtained in a viscosimeter. (The scatter over almost an order of
magnitude is not unusual for viscosity values.)

We repeated single experiments with PS films of 50~nm and 230~nm
thickness on both types of substrate and obtained identical values
for the viscosity and the slip length. Qualitative comparison of
the corresponding rims demonstrates that it is possible to induce
a transition between an oscillatory shape and a monotonic decay by
solely changing the initial film thickness $H$: as predicted, we
observe that thicker films tend to more pronounced oscillations,
whereas on thinner films oscillations are suppressed.

\begin{figure}
\centerline{
\includegraphics*[width=0.7\linewidth]{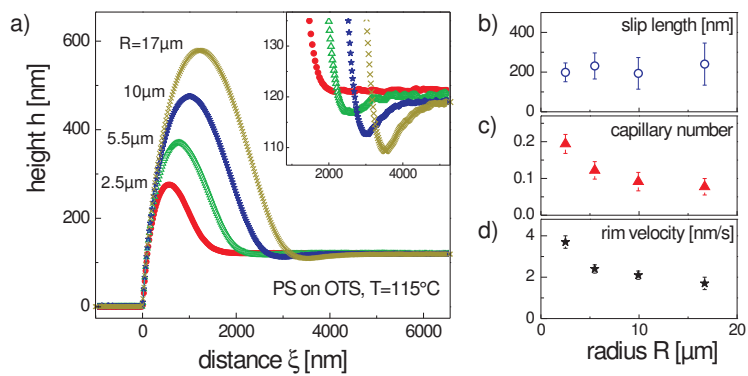}}
\caption{Rim analysis at different dewetting stages. a) Cross
section of in situ SPM scans of the rim around a hole at radii
2.5~$\mu$m (filled circles), 5.5~$\mu$m (open triangles),
10~$\mu$m (stars), and 17~$\mu$m (crosses) (i.e., at different
times) in a 130~nm thick PS(13.7k) film on OTS at 115$^\circ$C.
Between $R = 2.5$~$\mu$m and $R = 5.5$~$\mu$m we observe a
transition from monotonic to oscillatory rim shape (see the
enlarged view in the inset). b) Slip length and c) capillary
number determined from the profiles shown in a). While the
capillary number decreases by over a factor of two, the slip
length stays constant. d) Rim velocity as determined
independently.} \label{fig3}
\end{figure}

We also analysed rim profiles of holes at different dewetting
stages, as shown in fig.~\ref{fig3}. As the radius $R$ of a hole
grows, dewetting slows down due to the growing rim. As predicted,
a decrease in rim velocity results in a more pronounced
oscillatory shape. Thus, the rim changes its shape while growing
(see also ref.~\cite{Mue05}). Nevertheless, the slip length
extracted from these profiles remains constant within the error
bars, i.e., it is independent of the velocity (see
fig.~\ref{fig3}b). We also verified that the same is true for the
estimates of the viscosity obtained from the capillary number,
i.e., we found that the values for Ca that resulted from the fit
of the profiles are proportional to the rim velocity, cf.
fig.~\ref{fig3}c) and \ref{fig3}d).

\begin{figure}
\centerline{
\includegraphics*[width=0.9\linewidth]{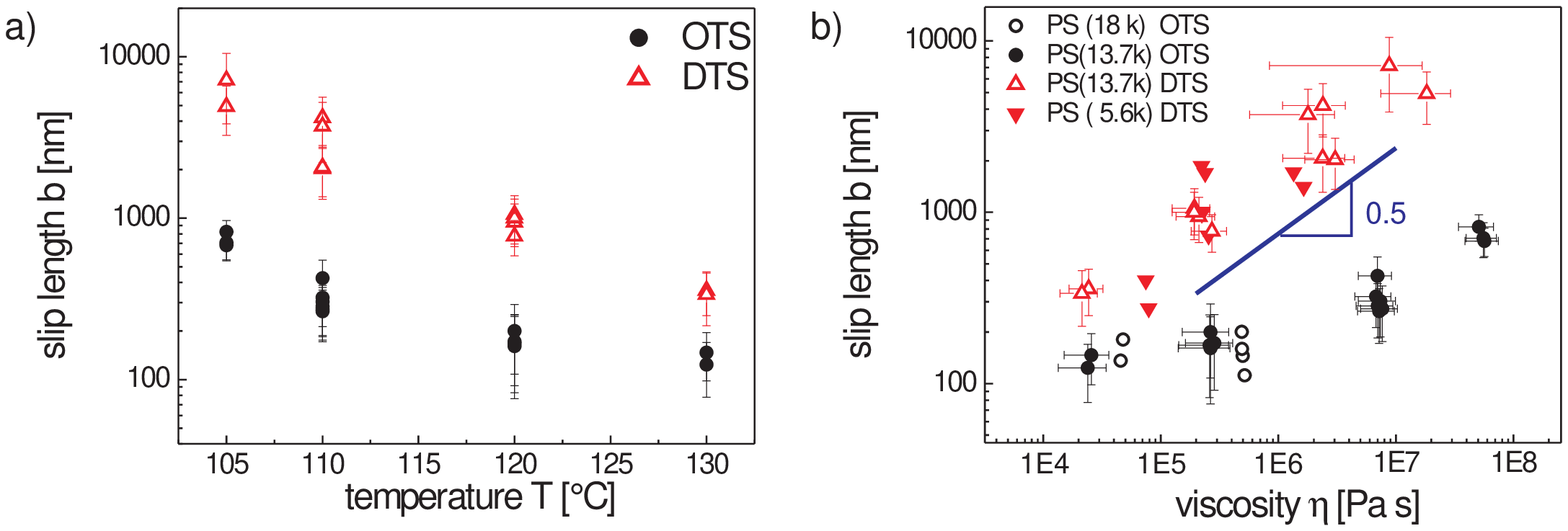}}
\caption{a) Slip length of polystyrene films on OTS and DTS as a
function of temperature. The values are determined via eq.
(\ref{eq_polynomial}) from the rim profiles of 130~nm thick
PS(13.7k) films. b) Slip length of all investigated films plotted
against their respective viscosity. Since the data collapse onto
two curves, one for OTS (circles) and one for DTS substrates
(triangles), $b$ can be written as function of $\eta$. The solid
line marks a slope of 0.5, indicating that the DTS data is better
fitted by $b \sim \eta^{1/2}$ rather than a linear dependence. The
OTS data does not follow a power law.} \label{fig4}
\end{figure}

Above we checked the consistency of our method and state that it
yields reliable values for the viscosity and the slip length. We
can now proceed and analyse the dependence of the slip length on
temperature and substrate type. The results are summarised in
fig.~\ref{fig4}a). We find that on OTS slippage is reduced by
about one order of magnitude as compared to DTS. This result is in
agreement with the observed dewetting velocities: Holes grow
significantly faster on the DTS brush. Additionally, on both types
of coating, the slip length decreases with increasing temperature,
as does the viscosity.

The question arises whether $b$ can be written as a function of
$\eta$, e.g., as $b = \eta /k$ with a temperature-independent
friction coefficient $k$. The Navier slip condition is often
written in terms of this proportionality factor $k$ between slip
velocity and shear stress at the substrate. To answer this, we
repeated some experiments with PS(5.61k) and PS(18k), henceforth
changing the viscosity at constant temperature. All samples had
chain lengths below the entanglement length in order to exclude
viscoelastic effects. As expected, the viscosity we gain from the
rim profiles increases with molecular weight. Plotting the slip
length of all samples versus the corresponding viscosity (see
fig.~\ref{fig4}b) we can collapse the data onto two master curves,
one for OTS and one for DTS. The dependence, however, is clearly
non-linear. Indeed, the DTS data is nicely fitted by a power law
$b \sim \eta^q$ with exponent $q=1/2$. This corresponds to a
friction coefficient $k$ that increases with viscosity. On OTS,
the data for high viscosity values can be fitted by the same power
law, but for small values of $\eta$  the slip length increases
only slightly with the viscosity.

\section{Summary and discussion}

We analysed the profiles of rims around holes in thin dewetting
Newtonian PS films on OTS and DTS covered silicon wafers. To
capture the form of the profile, we used a thin film model in the
strong slip regime. Fitting the theoretical functions to the
experimental data, we gained the capillary number and the slip
length. Additional knowledge of the dewetting velocity allowed the
extraction of the viscosity of the liquid via the capillary
number. Analysing profiles of films of different molecular weight,
we found that the slip length on the short brush (DTS) is
significantly larger than on the long brush (OTS), leading to
higher dewetting velocities on DTS as compared to OTS. We moreover
observed that the slip length is a non-linear monotonically
increasing function of the viscosity. On DTS, the slip length
increases roughly with the square root of the viscosity.
Furthermore, the data shown in fig.~\ref{fig4}b) indicate that the
chain length of the polymer molecules seems to play no significant
role for the amount of slippage. This is in accordance with the
expectation for polymer melts below the entanglement length.

Our finding, that the friction coefficient $k$ is a monotonically
increasing function of the viscosity indicates that the mechanism
for the momentum transfer between the liquid and the solid and the
microscopic origin of the viscosity are related. Since the
viscosity decreases with temperature, the friction coefficient
must also decrease with temperature. Recent molecular dynamics
simulations show that the anchoring of a polymer melt on a brush
decreases with decreasing penetration of the melt into the brush
\cite{Pas06}. Moreover, the penetration was found to be lower for
smaller melt density. Unfortunately, all simulations were
performed at the same temperature such that we can only speculate
about the influence of temperature on the anchoring. But since our
brushes are quite short and densely packed, we do not expect large
changes in the brush structure over the temperature range covered
in our experiments, but the melt density decreases slightly,
probably enough to explain our findings. However, this must remain
a speculation unless further corroborated by experiments and
simulations.

Another speculation concerns the origin of the different slip
lengths on the two types of brushes. The OTS and DTS brushes might
have slightly different properties, e.g., not exactly identical
density or short-ranged interaction forces, which give rise to a
different anchoring density of the polymer onto the brush. We
have, however, no insight in the molecular motion of the fluid
particles near the wall, since we analysed the film surfaces and
used continuum theory based on hydrodynamics. Hence, we cannot
give any prediction about the molecular mechanism that leads to
the observed slip length. Here we see a playground for further
experimental investigations, probably involving scattering
techniques to gain insight into the details of anchoring.

To conclude, we have developed a new consistent method to
determine the slip length as well as the viscosity of dewetting
thin films based on Newtonian hydrodynamics. This powerful tool
now offers the potential to quantify slippage in comprehensive
studies using various substrates and liquids. It is now possible
to systematically investigate the influence of system parameters,
e.g., substrate roughness, viscosity, and even viscoelasticity. To
explain the latter, we may note that our current model only
applies directly to Newtonian liquids. Highly viscous fluids, in
particular polymeric liquids, often exhibit viscoelastic
behaviour, i.e., the fluid has an internal time constant $\tau$
for the relaxation of stresses generated by shear. However, these
viscoelastic properties are only relevant if the Weissenberg
number $Wi = \tau \dot{\gamma}$, i.e., the product between the
relaxation time and the shear rate is of order one or larger. Thus
our Newtonian model remains valid for viscoelastic fluids, if they
move slowly enough. A model that captures viscoelastic effects is
currently under way.

\acknowledgments
The authors thank Marcus Müller for fruitful
discussions. This work was supported in part by the
Heisenberg-scholarship DFG Grant MU 1626/3 (AM), the DFG Research
Center \textsc{Matheon} Berlin (AM and BW), and the Grant Ja 905/3
within the priority program 1164 (RF and KJ). RF and KJ
acknowledge generous support of Si wafers by Siltronic AG,
Burghausen, Germany.

\end{document}